# Electronic structure and Fermi surface of new K intercalated iron selenide superconductor $K_xFe_2Se_2$

I.R. Shein, * A.L. Ivanovskii

*Institute of Solid State Chemistry, Ural Branch of the Russian Academy of Sciences, 620990, Ekaterinburg, Russia*

A B S T R A C T

Using the *ab initio* FLAPW-GGA method we examine the electronic band structure, densities of states, and the Fermi surface topology for a very recently synthesized $ThCr_2Si_2$-type potassium intercalated iron selenide superconductor $K_xFe_2Se_2$. We found that the electronic state of the *stoichiometric* $KFe_2Se_2$ is far from that of the isostructural iron pnictide superconductors. Thus the main factor responsible for experimentally observed superconductivity for this material is the deficiency of potassium, *i.e.* the hole doping effect. On the other hand, based on the results obtained, we conclude that the tuning of the electronic system of the new $K_xFe_2Se_2$ superconductor in the presence of K vacancies is achieved by joint effect owing to structural relaxations and hole doping, where the structural factor is responsible for the modification of the band topology, whereas the doping level determines their filling.

* Corresponding author.
*E-mail address:* shein@ihim.uran.ru

# 1. Introduction

Among the recently discovered [1] iron-based high-temperature superconductors (SCs) two main groups are found: so-called iron pnictide (Fe*Pn*, where *Pn* are P, As, and Bi) and iron chalcogenide (Fe*Ch*, where *Ch* are S, Se, and Te) systems. For Fe*Pn* SCs, several structural families are known: ternary 111 (such as LiFeAs), 122 (such as $BaFe_2As_2$), quaternary 1111 (such as LaFeAsO), and five-component 32225 (such as $Sr_3Sc_2Fe_2As_2O_5$) and 42226 (such as $Sr_4V_2Fe_2As_2O_6$) systems. Their layered crystal structures include $[Fe_2Pn_2]$ building blocks alternating with planar sheets of alkaline or alkaline-earth metals (for 111 or 122 phases) or with more complex structural blocks (for 1111, 32225, and 42226 phases), reviews [2-9].

In turn, the Fe*Ch* SCs form a family of simplest binary layered materials with blocks, in which Fe cations are tetrahedrally coordinated with *Ch* ions. These systems are known also as 11 phases. The transition temperatures $T_C$ for these materials (for example, for FeSe) at ambient conditions are quite low and do not exceed 8K. Considerable efforts were undertaken recently to improve $T_C$ of these binary 11 Fe*Ch* phases: by partial substitutions of chalcogens (S ↔ Se ↔ Te), by substitutions of transition metals on the iron site, by external pressure etc., see reviews [10,11].

Very recently, two new unique ternary Fe*Ch* SCs with enhanced $T_C$ were discovered by intercalation of alkaline metal ions between $[Fe_2Se_2]$ blocks. Their nominal compositions are $K_{0.8}Fe_2Se_2$ [12] and $Cs_{0.8}Fe_2Se_{1.96}$ [13]. These phases possess the highest $T_C \sim 30K$ [12] and $T_C \sim 27$ K [13] among all known Fe*Ch* SCs under ambient pressure, see reviews [10,11]. Moreover, like the family of 122 Fe*Pn* materials, these phases adopt a tetragonal structure of the $ThCr_2Si_2$ type (space group *I*/4*mmm*).

According to [12,13], the enhanced superconductivity of the novel 122 Fe*Ch* phases may be related to structural indicators, such as the so-called anion height (Δz, the distance of *Pn* (*Ch*) atoms from the Fe plane inside $[Fe_2Pn(Ch)_2]$ blocks) and Se-Fe-Se bond angles, which for these materials are closer to their "optimum" values (~1.38 Å and 109.47°) for iron based SCs, see [2,3,14]. The other key factor can be related to K(Cs) non-stoichiometry, which is responsible for the electronic concentration in the systems and therefore for the type of band filling. However, no data about the electronic properties of these novel 122 Fe*Ch* phases are hitherto available.

In view of these circumstances, in this Communication we present a detailed *ab initio* study of the recently discovered $ThCr_2Si_2$-type $KFe_2Se_2$ and focus our attention on its electronic properties and the Fermi surface (FS) topology – as dependent on the above mentioned structural and electronic factors.

**2. Models and computational aspects**

According to [12], the new potassium intercalated iron selenide $K_xFe_2Se_2$ adopts a tetragonal $ThCr_2Si_2$-type structure (space group I4/mmm; #139). The atomic positions are K: 2*a* (0, 0, 0), Fe: 4*d* (0, ½, ¼) and Se: 4*e* (0, 0, $z_{Se}$), where $z_{Se}$ is the so-called internal coordinate. Further we shall examine this material with the nominal composition $KFe_2Se_2$.

Firstly, we focused our attention on the electronic properties of $KFe_2Se_2$ as depending on the structural factors. For this purpose, full structural optimization of the stoichiometric $KFe_2Se_2$ was performed both over the lattice parameters and the atomic positions including the internal coordinate $z_{Se}$ as no detailed atomic coordinates were known for this phase, and then the electronic properties of this phase were calculated. Next, similar calculations were continued using the experimental structural parameters [12] for the composition $K_{0.8}Fe_2Se_2$. For these two systems abbreviated further as $KFS^{calc}$ and $KFS^{exp}$, their electronic band structures and FSs were obtained and analyzed.

Secondly, the effect of band filling on the electronic properties and the Fermi surface topology for $K_xFe_2Se_2$ was considered. Here, using the rigid-band model, we removed 0.4*e*, 0.3*e*, and 0.2*e* from the $KFS^{calc}$ and $KFS^{exp}$ systems and in this way simulated K-deficient compositions $K_xFe_2Se_2$ with x = 0.6, 0.7, and 0.8, respectively.

All our calculations were carried out by means of the full-potential method within mixed basis APW+lo (LAPW) implemented in the WIEN2k suite of programs [15]. The generalized gradient correction (GGA) to exchange-correlation potential in the PBE form [16] was used. The plane-wave expansion was taken to $R_{MT} \times K_{MAX}$ equal to 8, and the $k$ sampling with 12×12×12 $k$-points in the Brillouin zone was used. The hybridization effects were analyzed using the densities of states (DOSs), which were obtained by a modified tetrahedron method [17].

## 3. Results and discussion

The experimental [12] and calculated structural parameters for $KFS^{exp}$ and $KFS^{calc}$ are summarized in the Table 1. $KFS^{calc}$ may be viewed as a "compressed" phase with reduced $a$ and $c$ parameters as compared with $KFS^{exp}$.

The calculated band structures, densities of states (DOSs), and Fermi surfaces for $KFS^{exp}$ and $KFS^{calc}$ are depicted in Figs. 1-3. For both systems, their electronic spectra are similar (see Fig. 1), namely (i) the Se $4p$ states occur between -6.5 eV and -3.5 eV with respect to the Fermi level ($E_F$= 0 eV); (ii) the bands between -2.4 eV and $E_F$ are mainly of the Fe $3d$ character, and (iii) the contributions from the valence states of K to the occupied bands are very small. Thus, as with related 122 Fe*Pn* phases [2-7], potassium atoms in $KFe_2Se_2$ are in the form of cations close to $K^{1+}$ and provide charge transfer into conducting blocks $[Fe_2Se_2]^{n-}$.

Let us focus on the most important features of the electronic band structure, namely on the low-dispersive bands, which, for the family of 122 Fe*Pn* materials, cross the Fermi level and are responsible for the features of the FS topology [2,3,9,18,19]. The examined $KFe_2Se_2$ phase has an increased number of valence electrons (*nve* = 29 *e* per formula unit) as compared with FeAs-based materials (for example, $BaFe_2As_2$, where *nve* = 28 *e* per formula unit). This leads to shifting of the Fermi level to the upper Fe $d$-like bands with higher $k_z$ dispersion. As a result, both for KFS$^{exp}$ and KFS$^{calc}$ the above low-dispersive $d_{xy,\ xz+yz}$ bands are below $E_F$, Fig. 2.

It is well known that the Fermi surface of ThCr$_2$Si$_2$-like 122 Fe$Pn$ SCs adopts a characteristic quasi-two-dimensional (2D) multiple sheet topology and consists of cylinder-like hole pockets along the $\Gamma$-$Z$ direction and cylinder-like electron pockets along the $X$-$P$ direction , see [2,3,9,18,19]. The above shift of the Fermi level leads to serious differences in the FSs for KFe$_2$Se$_2$ *versus* 122 Fe$Pn$ SCs (see Fig. 3): both for KFS$^{exp}$ and KFS$^{calc}$ the Fermi surfaces contain closed hole-like pockets centered at $Z$ point - instead of cylinder-like sheets for 122 Fe$Pn$ materials.

On the other hand, we can see (Fig. 2) that for KFS$^{exp}$ with increased lattice parameters the low-dispersive $d_{xy,\ xz+yz}$ bands come much nearer to $E_F$, and the closed hole-like pockets (around $Z$) are extended along the $k_z$ direction acquiring a pronounced conical shape. Additionally this shift is accompanied by growth of total DOS at the Fermi level, $N(E_F)$, from 2.57 states/eV · f.u. for KFS$^{calc}$ up to 3.81 states/eV · f.u for KFS$^{exp}$, see Table 2. However based on the results obtained we find that the electronic state of the *stoichiometric* KFe$_2$Se$_2$ is far from those required for the iron-based high-temperature superconductors.

Thus, it may be concluded that the main factor responsible for the experimentally observed superconductivity for this material is the deficiency of potassium, *i.e.* the hole doping effect. Indeed, it is possible to explain this effect qualitatively using the band picture for the *stoichiometric* KFe$_2$Se$_2$, Fig. 2. If K$^{1+}$ ions are partially removed, the *nve* in the system lowers, high-dispersive Fe $d$-like bands get empty, and the Fermi level shifts to low-dispersive $d_{xy,\ xz+yz}$ bands forming an electronic picture typical of 122 Fe$Pn$ SCs.

To confirm this tendency, in Fig. 3 we present the Fermi surfaces for a set of potassium-deficient phases K$_x$Fe$_2$Se$_2$ for x = 0.8, 0.7, and 0.6 as calculated in the framework of the above rigid-band model both for KFS$^{exp}$ and KFS$^{calc}$ systems.

As is seen, for KFS$^{calc}$, which keeps the optimized geometry for *stoichiometric* KFe$_2$Se$_2$, at the first stage (from x= 1.0 to x = 0.7) the size of the closed Z-centered pocket decreases, and only at a high level of K deficiency (x = 0.6) a cylinder-like hole sheet arises.

On the contrary, for KFS$^{exp}$ (with experimental lattice parameters for K$_{0.8}$Fe$_2$Se$_2$ phase) the cylinder-like sheet is formed already at x = 0.8, and further growth of K deficiency leads only to an increase in the diameter of this hole-like cylinder.

Thus, we conclude that the tuning of the electronic system of the new K intercalated iron selenide superconductor in the presence of potassium vacancies is achieved by joint effect owing to structural relaxations and hole doping. It is possible to assert that the structural factor is responsible for the modification of the band topology, whereas the doping level determines their filling.

## 4. Conclusions

In conclusion, we used the first-principle FLAPW-GGA approach to examine the band structure, density of states, and the Fermi surface topology for the recently synthesized ThCr$_2$Si$_2$-type K intercalated iron selenide superconductor K$_x$Fe$_2$Se$_2$.

Our results show that the electronic state of the *stoichiometric* KFe$_2$Se$_2$ is far from that of isostructural iron pnictide superconductors, and the main factor responsible for experimentally observed superconductivity for this material is the deficiency of potassium, *i.e.* the hole doping effect. On the other hand, taking into consideration the results obtained, we conclude that the tuning of the electronic system of the new K$_x$Fe$_2$Se$_2$ superconductor in the presence of potassium vacancies is achieved by joint effect owing to structural relaxations and hole doping, where the structural factor is responsible for the modification of the band topology, whereas the doping level determines their filling.

**Acknowledgments**

Financial support from the RFBR (Grants 09-03-00946 and 10-03-96008) is gratefully acknowledged.

**Table 1.**
Structural parameters for $KFe_2Se_2$: lattice constants ($a, c$, Å), internal coordinate ($z_{Se}$), bond length ($d$, Å), bond angles ($\Theta$,°), and anion height ($\Delta z$, Å).

| System * / parameter | KFS$^{calc}$ | KFS$^{exp}$ [12] |
|---|---|---|
| $a$ | 3.8608 | 3.9136 |
| $c$ | 13.8369 | 14.0367 |
| $z_{Se}$ | 0.3452 | 0.3539 |
| $d$(K-Se) | 3.4700× 8 | 3.4443 × 8 |
| $d$(Fe-Se) | 2.3370× 4 | 2.4406 × 4 |
| $d$(Fe-Fe) | 2.7299× 4 | 2.7673 × 4 |
| $\Theta$ | 111.41× 4; 108.52× 2 | 110.93× 4; 106.60× 2 |
| $\Delta z$ | 1.3165 | 1.4237 |

* see text

**Table 2.**
Total and partial densities of states ($N(E_F)$, states/eV·f.u.) at the Fermi level for $KFe_2Se_2$.

| System / $N(E_F)$ | KFS$^{calc}$ | KFS$^{exp}$ | System / $N(E_F)$ | KFS$^{calc}$ | KFS$^{exp}$ |
|---|---|---|---|---|---|
| total | 2.571 | 3.811 | Fe $3d_{xz+yz}$ | 1.052 | 1.711 |
| Fe $3d$ | 1.772 | 2.873 | Se $4s$ | 0.008 | 0.011 |
| Fe $3d_{z^2}$ | 0.023 | 0.032 | Se $4p$ | 0.122 | 0.143 |
| Fe $3d_{xy}$ | 0.631 | 1.058 | Se $3d$ | 0.031 | 0.030 |
| Fe $3d_{x^2-y^2}$ | 0.067 | 0.072 | | | |

**FIGURES**

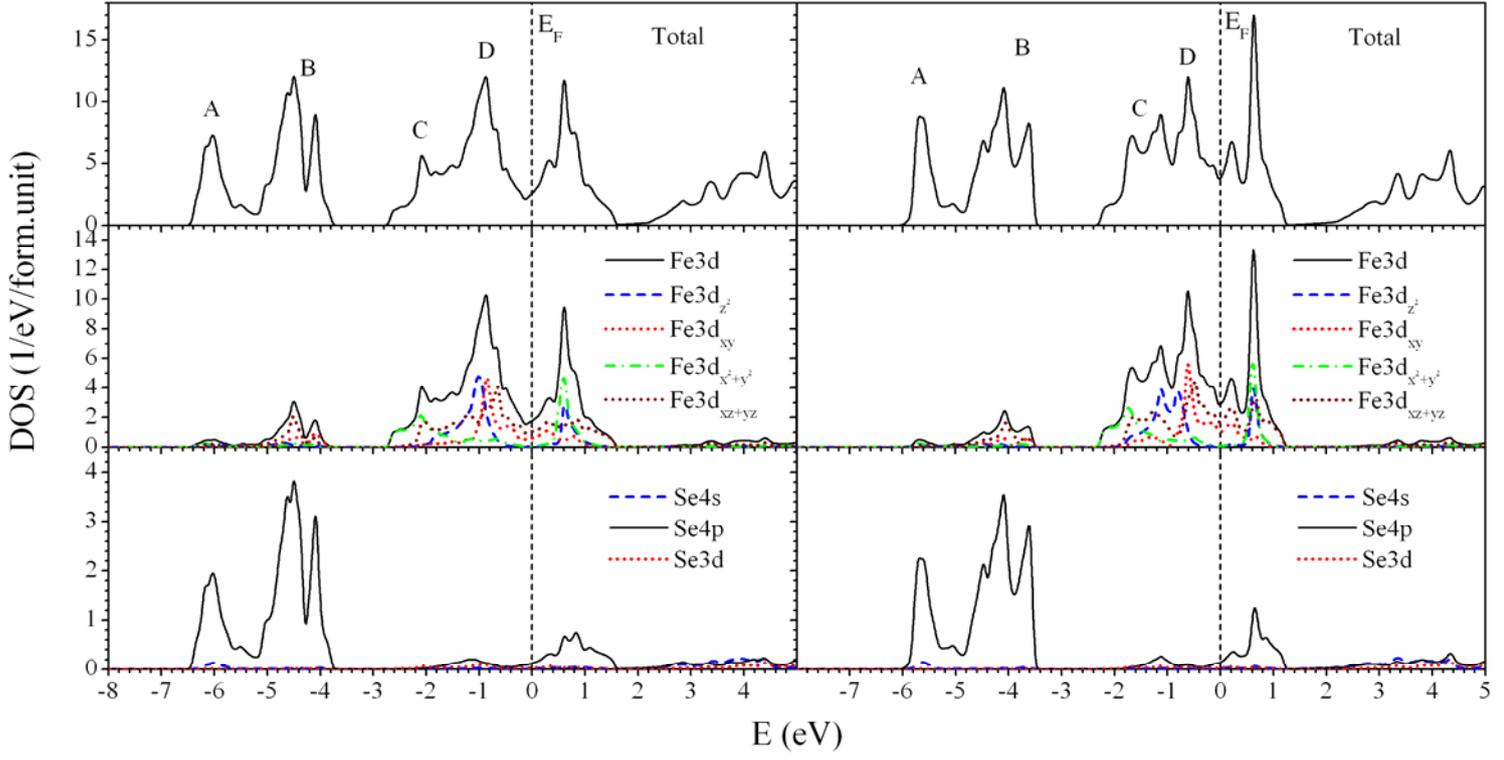

**Fig. 1.** Total (*upper panels*) and partial densities of states for KFS$^{calc}$ (left), and for KFS$^{exp}$ (right).

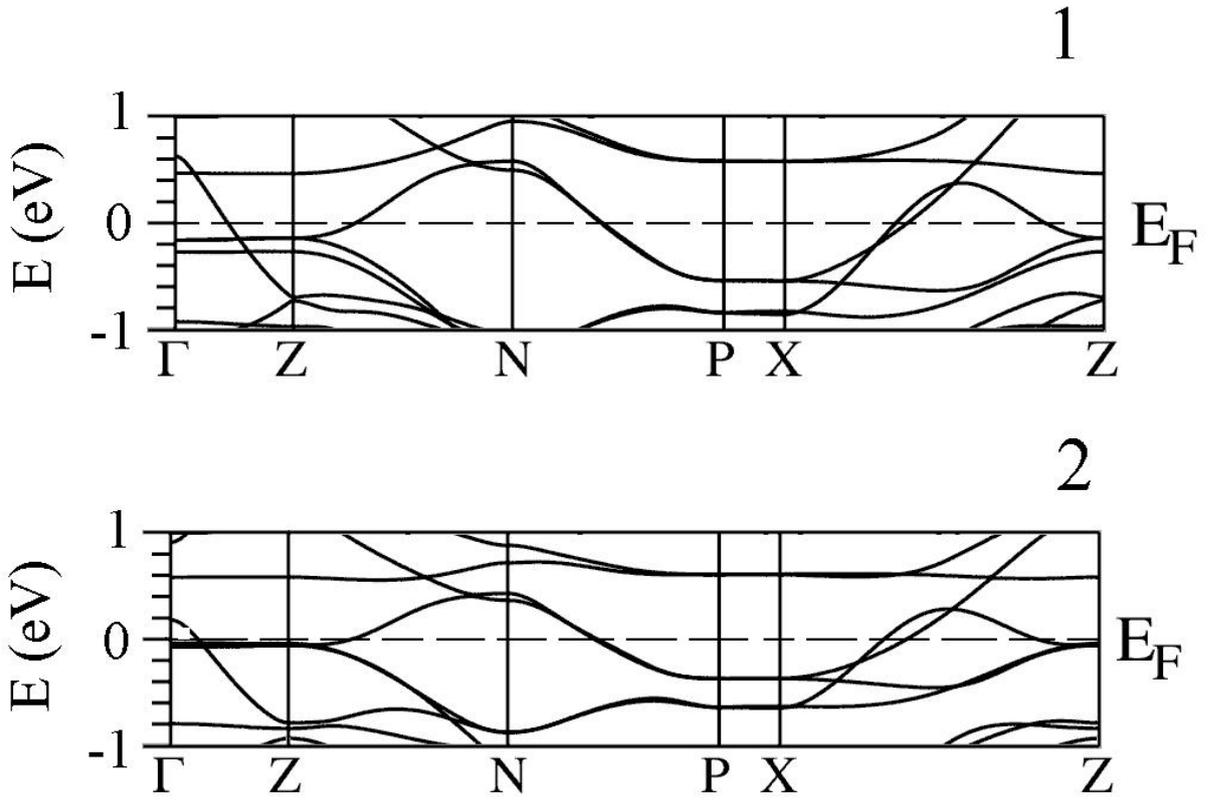

**Fig. 2.** Near-Fermi electronic bands for KFS$^{calc}$ (1) and for KFS$^{exp}$ (2).

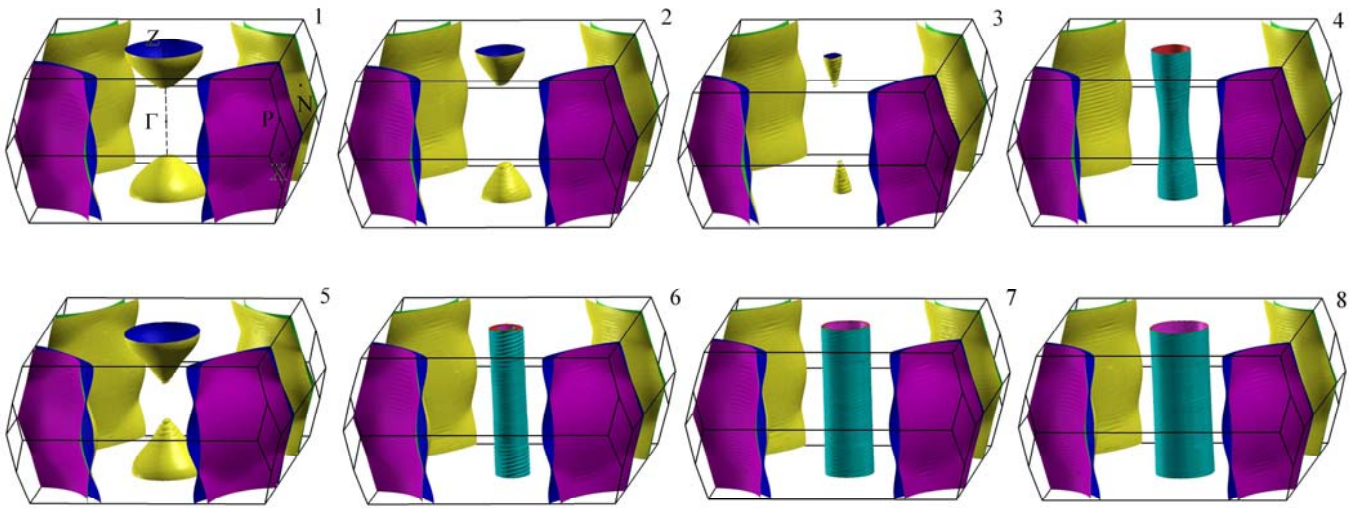

**Fig. 3.** Fermi surfaces for KFS$^{calc}$ for x = 1.0 (1), 0.8 (2), 0.7 (3), and 0.6 (4) (*top panel*) and for KFS$^{exp}$ for x = 1.0 (5), 0.8 (6), 0.7 (7), and 0.6 (8) (*bottom panel*).